\documentclass[preprint2]{aastex}

\usepackage{natbib,graphicx,color} \singlespace

\shorttitle{Shrinking of Galaxies in Clusters}
\shortauthors{Cypriano et al.}

\newcommand\eq{\begin{equation}}
\newcommand\eeq{\end{equation}}
\newcommand\eqn{\begin{eqnarray}}
\newcommand\eeqn{\end{eqnarray}}
\newbox\grsign \setbox\grsign=\hbox{$>$} \newdimen\grdimen
\grdimen=\ht\grsign
\newbox\simlessbox \newbox\simgreatbox
\setbox\simgreatbox=\hbox{\raise.5ex\hbox{$>$}\llap
     {\lower.5ex\hbox{$\sim$}}}\ht1=\grdimen\dp1=0pt
\setbox\simlessbox=\hbox{\raise.5ex\hbox{$<$}\llap
     {\lower.5ex\hbox{$\sim$}}}\ht2=\grdimen\dp2=0pt

\newbox\simppropto
\setbox\simppropto=\hbox{\raise.5ex\hbox{$\sim$}\llap
     {\lower.5ex\hbox{$\propto$}}}\ht2=\grdimen\dp2=0pt

\begin{document}

\title{Shrinking of Cluster Ellipticals: a Tidal Stripping explanation and
Implications for the Intra-Cluster Light.}

\title{Shrinking of Cluster Ellipticals}

\author{Eduardo S. Cypriano}
\affil{Southern Astrophysics Research Telescope, Casilla 603, La Serena, Chile
and Laborat\'orio Nacional de Astrof\'{\i}sica, 
CP 21, 37500-000 Itajub\'a - MG, Brazil}
\email{ecypriano@ctio.noao.edu}

\author{Laerte Sodr\'e Jr.}
\affil{Departamento de Astronomia, Instituto de Astronomia, Geof\'{\i}sica
e Ci\^encias Atmosf\'ericas da USP, Rua do Mat\~ao 1226, Cidade
Universit\'aria, 05508-090, S\~ao Paulo, Brazil}
\email{laerte@astro.iag.usp.br}

\author{Luis E. Campusano \altaffilmark{1}}
\affil{Observatorio Astron\'omico Cerro Cal\'an, Departamento de 
Astronom\'{\i}a, Universidad de Chile, Casilla 36-D, Santiago, Chile}
\email{luis@das.uchile.cl}

\author{Daniel A. Dale \altaffilmark{1}}
\affil{Department of Physics and Astronomy, University of Wyoming, Laramie, WY 82071}
\email{ddad@uwyo.edu}


\and
\author{Eduardo Hardy \altaffilmark{1,2}}
\affil{National Radio Astronomy Observatory, Casilla El Golf 16-10, Chile}
\email{ehardy@nrao.edu}

\altaffiltext{1}{Visiting Astronomer at the Cerro Tololo Inter-American 
Observatory, National Optical Astronomy Observatories, which is operated by the
Association of Universities for Research in Astronomy, Inc., under
a cooperative agreement with the National Science Foundation.}
\altaffiltext{2}{The National Radio Astronomy Observatory is a facility of
the National Science Foundation operated under cooperative agreement by
Associated Universities, Inc.}

\begin{abstract}
{We look for evidence of tidal stripping in elliptical galaxies through the
analysis of homogeneous CCD data corresponding to a sample of 228 elliptical 
galaxies belonging to 24 clusters of galaxies at $0.015<z<0.080$. We investigate
departures from the standard magnitude-isophotal size relation, as a function
of environmental (cluster-centric distance, local galaxy density) and
structural (cluster velocity dispersion, Bautz-Morgan type) properties. We find
that, for any particular galaxy luminosity, the ellipticals in the inner and
denser regions of the clusters are about 5\% smaller than those in the outer
regions, which is in good agreement with the finding of \citet{2S}
based on photographic photometry. The null hypothesis (ie., galaxy sizes
are independent of the cluster-centric distance or density) is rejected at a
significance level of better than 99.7\%.
Numericals models of \citet{AW2} predict that tidal stripping can
lead to changes in the whole structure of ellipticals producing shrinkage and
brightening of the galaxy, qualitatively consistent with our measurements and
also with the findings of \citet{trujillo}, that more centrally concentrated
ellipticals populate denser regions. Our observational results can be
interpreted as evidence for stripping of stars from ellipticals in the
central/denser regions of clusters, contributing to the intra-cluster light
observed in these structures.}
\end{abstract}


\section{Introduction}
The existence of diffuse optical light in the central region of galaxy clusters
is well known \citep{z51, dV60}, but its  very low surface
brightness precluded quantitative analysis.

More recently, due to advances in observational and data reduction  techniques,
accurate measurements of the total amount of this light and its color and
spatial distribution became feasible  for rich clusters \citep{Uson,
Scheick, Gonzalez,Gonzalez2005,Feldmeier2,Feld2004a}. Moreover, single objects, like
planetary nebulae, red-giant stars or supernovae  have been detected in the
intra-cluster medium (ICM)  \citep{T&W, Arnaboldi,A.02,A.03, Feldmeier,
Ferguson,Feld2004b}.

These studies indicate that between $\sim$10\%
to 20\% (or higher) of the total stellar  light of the clusters comes from
stars in the ICM. These stars have, probably, been stripped from  cluster
galaxies after galaxy-galaxy \citep{GO,richstone} or galaxy-cluster 
interactions \citep{Merritt84}.  As a galaxy moves through the cluster, it is
subject to the gravitational forces of their neighbors, as well as of the
cluster   as a whole. As a result of the action of these tidal forces, some
galaxy stars  may be accelerated to velocities larger than the local escape
velocity, being removed from the parent galaxy.  This process is called  {\it
tidal stripping}.  In some cases the tidal forces may be so strong that the
galaxy as a whole is disrupted. Indeed, a significant fraction of galaxy stars
may be dispersed in the ICM after disruption of dwarf spheroidals or low 
surface brightness disk galaxies \citep[e.g.][]{Omar}, as suggested by the
light plumes observed in clusters like Centaurus and Coma \citep{Roldan,
Korchagin}. 

In the present work we investigate whether stars stripped of ellipticals can
make a significant contribution to the intra-cluster light (ICL). There are some
hints that point towards this direction. Firstly, ellipticals are basically
found in the central, denser, cluster regions, where close encounters between
galaxies are very common. Their almost radial orbits \citep{Amelia} lead them
to several close approximations  to the cluster center, where tidal  forces
are very strong.  The diffuse light is often distributed  like a halo around
the brightest cluster galaxy (BCG), extending  sometimes beyond $\sim 500$ kpc.
\citet{Gonzalez2005} in their study  of the surface brightness distribution of BCGs
in 24 clusters identify a component, independent of the BCG, that is possibly
due to intracluster stars and dynamically linked to the cluster as a whole.

The effects of galaxy-galaxy tidal collisions were studied by \citet{AW1,AW2}
through the use of the impulsive approximation and N-body numerical simulations
of close collisions between galaxies with roughly the same mass and with mass
distributions consistent with de Vaucouleurs and King profiles.  They
quantified the mass and energy changes during such collisions and studied the
changes of the mass profiles. They found that the $r^{1/4}$ profile is robust,
since it is recovered after a collision, but with different parameters. After
strong encounters the central surface brightness becomes brighter and the
effective and isophotal radius decrease as mass is stripped out. More recently,
the high-resolution N-body+SPH  numerical simulations of rich clusters by
\citet{willman04} and \citet{murante} indicate that the accumulation of ICL  
is an ongoing process, linked to infall and stripping events, and that the stars
composing the ICL are on average older than the stars in cluster galaxies.


Other works have considered the effects of a smooth cluster potential on a galaxy.
Using the impulsive approximation \citep{Merritt84,Mamon}, or high resolution
numerical simulations \citep{Ghigna}, it can be shown that cluster galaxies are
truncated at the tidal radius:

\begin{equation}
\label{rmare}
r_{tid} = C \left({\sigma_g \over \sigma_{cl}}\right) r_{p},
\end{equation}

\noindent where $\sigma_g$ and $\sigma_{cl}$ are the velocity dispersions 
of the galaxy and the cluster, respectively, $r_{p}$ is the radius of
maximum approximation between a galaxy and the cluster center, and $C$ is a 
non-dimensional constant that depends of several factors, like galaxy 
orbits, and is close to unity. This process is often called {\it tidal
truncation}.

It is not clear yet what process is more effective in removing stars 
from cluster galaxies. Tidal truncation is indeed more efficient in removing
halo stars, since $r_{tid}$ is larger than the optical radius 
for the majority of the galaxies. Anyway, as pointed out by \citet{Gnedin},
mass is not only lost by instantaneous stripping after cluster-galaxy shocks,
but also by secular tidal heating.
   
The first observational detection of tidal stripping of stars from elliptical
galaxies came from \citet{2Sa,2Sb,2Sc,2S}, with a sample of 400 ellipticals and
S0's located in 6 different clusters, deeply imaged with photographic plates.
These authors found that ellipticals in the center of rich clusters have
isophotal radii 10-30\% smaller at a given magnitude than ellipticals in poor
clusters or in  the outskirts of rich ones, estimating that up to $\sim$ 37\%
of the luminous mass of the stripped ellipticals have been lost to the ICM.

Despite of the great impact of this work, it has been severely criticized,
mainly because it was based on photographic photometry
\citep[e.g.][]{Dressler}.
\citet{Giuricin} questioned the statistical analysis done by the Strom \& Strom, and 
using  photo-electric and CCD data of $\sim 160$ elliptical galaxies from
the sample of
\citet{Burstein}, did not find any dependence of the magnitude-size relation
properties of  elliptical galaxies with the environment.
In this paper we will revisit this topic, using CCD data homogeneously 
collected and reduced.

The layout of the paper is as follows. In Section 2 we describe the
observational material and the cluster sample, we present our method for the
extraction of galaxy photometric parameters, as well as the selection criteria
of the final galaxy sample. In  Section 3 we present our results on galaxy
sizes, that are discussed in Section 4. In Section 5 we summarize the main
conclusions of this work.  Throughout this paper, we adopt, when necessary,  
a $\Lambda$CDM cosmology with
$H_0 = 70$ h$_{70}$ km s$^{-1}$ Mpc$^{-1}$, $\Omega_M$ = 0.3 and
$\Omega_\Lambda$ = 0.7. 
 
\section{The data}
 
\subsection{The cluster sample and the observations}
\label{obs}

The galaxy sample discussed in this work was extracted from the set originally
obtained as part of  the Ph.D.  thesis of Daniel A. Dale on peculiar motions 
of clusters with $z < 0.1$ \citep{Dale97,Dale98,Dale99a,Dale99b,Dale99c}.
The original cluster sample was selected from nearby ($cz < 18000$ 
km s$^{-1}$) objects of the \citet{ACO} cluster catalogue, in order to 
cover as much and uniformly  the whole sky area as possible.
Although the cluster selection criteria and imaging characteristics were 
chosen to optimize the peculiar motion study, the resulting database
is useful for many other studies, since it was homogeneously observed and
reduced, has high photometric accuracy, and the cluster sample is 
representative of the low redshift cluster population \citep{cyp01}.
For the present study, only the Southern part of this sample 
has been used.
The clusters, and some of their properties that are relevant for this work, 
are presented in Table \ref{sampletab}. 

\begin{deluxetable}{lcccccc}
\tablewidth{0pt}
\tablecaption{Cluster sample \label{sampletab}}
\tablehead{
\colhead{Name} & \colhead{z}\tablenotemark{a} &
\colhead{$\sigma_{cl}$}\tablenotemark{a} & \colhead{B-M Type} &
\colhead{Total area} & \colhead{Fraction of area} & \colhead{N}\\
\colhead{} & \colhead{} & \colhead{(km s$^{-1}$)} & \colhead{}  &
\colhead{(square degree)} &  \colhead{(\%)} & \\
\colhead{(1)} & \colhead{(2)} & \colhead{(3)} & \colhead{(4)} & \colhead{(5)} & 
\colhead{(6)} & \colhead{(7)} }
\startdata
A85   & 0.0555  & ~941  & I      & 0.05 &   4.8 & 12\\
A114  & 0.0582  & ~911  & II     & 0.39 &  44.8 & 15\\
A119  & 0.0449  & ~685  & II-III & 0.05 &   5.3 &  6\\
A496  & 0.0326  & ~686  & I      & 0.28 &  16.6 & 19\\
A2670 & 0.0766  & ~873  & I-II   & 0.05 &  11.4 &  7\\
A2806 & 0.0265  & ~433  & I-II   & 0.26 &  34.0 &  4\\
A2877 & 0.0251  & 1026  & I      & 0.45 &  10.2 &  7\\
A2911 & 0.0800  & ~546  & I-II   & 0.32 &  12.8 &  2\\
A3193 & 0.0346  & ~638  & I      & 0.23 &  21.3 &  5\\
A3266 & 0.0586  & 1131  & I-II   & 0.18 &  15.0 & 17\\
A3376 & 0.0456  & ~641  & I      & 0.10 &  14.2 &  8\\
A3381 & 0.0372  & ~414  & I      & 0.09 &  39.3 &  2\\
A3389 & 0.0259  & ~511  & II-III & 0.10 &   5.8 &  1\\
A3395 & 0.0496  & ~823  & II     & 0.10 &   9.9 & 10\\
A3407 & 0.0415  & ~504  & I      & 0.13 &  29.3 &  3\\
A3408 & 0.0405  & ~900  & I-II   & 0.20 &  12.4 &  4\\
A3528 & 0.0538  & ~955  & II     & 0.31 &  26.7 & 16\\
A3558 & 0.0473  & ~935  & I      & 0.56 &  41.4 & 35\\
A3571 & 0.0384  & ~969  & I      & 0.05 &   2.1 &  6\\
A3574 & 0.0152  & ~447  & I      & 0.09 &   2.2 &  1\\
A3656 & 0.0195  & ~366  & I-II   & 0.19 &  15.5 &  5\\
A3667 & 0.0549  & ~987  & I-II   & 0.44 &  33.9 & 19\\
A3716 & 0.0454  & ~842  & I-II   & 0.40 &  33.4 & 13\\
A3744 & 0.0386  & ~624  & II-III & 0.31 &  37.9 &  7\\
A4038 & 0.0302  & ~826  & III    & 0.33 &  11.3 &  4\\
\enddata
\tablenotetext{a}{We got redshifts and velocity dispertions  form  Abell
cluster redshift compilation of Andernach \& Tago (2005, in prep.), see
\citet{Heinz} for a description.}                                                                                  

\tablecomments{
(1) name of the cluster in the Abell catalog \citep{ACO}; 
(2) redshift; 
(3) radial velocity dispersion;
(4) Bautz-Morgan Type \citep{BM};
(5) total imaged area;
(6) The ratio of the area covered by the imaging over the whole 
 area inside the $R_{200}$ radius ;
(7) number of ellipticals in the sample.
}                                                 
\end{deluxetable}

The observational material consists of several Kron-Cousins I band images
obtained with the 0.9m CTIO telescope. The details of the observations are
discussed elsewhere \citep{Dale97,Dale98} and here we only summarize them. The
detector used was the 2k$\times$2k Tek2k No.3 CCD, with a scale of 0.4 arcsec
per pixel, resulting in a field of 13.5$^\prime \times 13.5^\prime$ per
image. The exposure time was 600 seconds in all cases. The images reach 23.0
to 23.8 I mag arcsec$^{-2}$ at the 1 $\sigma$ level over the background, with a
median of 23.6 I mag arcsec$^{-2}$. The median seeing of the images is
1.52$\arcsec$ and ranges from 1.16$\arcsec$ to 2.80$\arcsec$.

These images, consisting of several pointings per cluster, in general do
not uniformly cover a cluster, since they were taken in regions near spiral
galaxies, although the central region of the clusters is always covered. 
For each cluster, the fraction of the projected area inside 
the $R_{200}$ radius (see section \ref{sample}) covered by the imaging is
also shown in Table \ref{sampletab}. The actual distribution of pointings
is presented in \citet{Dale97,Dale98,Dale99b}

\subsection{Determination of photometric parameters}
Most of the images used here were obtained under good photometric conditions.
The absolute calibration of the magnitudes was done by using 
standard photometric stars of \citet{Landolt}. The photometric  zero-point
calibration could be determined with a median accuracy of 0.018 mag, 
never larger than 0.031 mag.

In this work we have used an isophotal radius as an estimator  of the galaxy
size because its logarithm is very well correlated with the magnitudes
(Spearman rank correlation coefficient $r_s =  -0.99$). On the other hand, the
log of the radius that encloses half of a galaxy light (the effective radius  
${r_{e}}$) shows  a significantly poorer correlation with magnitudes  ($r_s = 
-0.86$). The Petrosian radius \citep{Petrosian}  is also less well-correlated
with magnitudes ($r_s =  -0.84$), which is is not unexpected, since elliptical
galaxies spanning a range of 5 absolute magnitudes have different profiles
\citep{Caon}.

The isophotal quantities are defined at the same  isophotal limit in the galaxy
rest frame ($\mu_{lim}$).  The corresponding isophotal threshold ($\mu_t$), in
the observer rest frame, is given by the following expression:

\begin{equation}
\label{mut}
\mu_t = \mu_{lim} + A_I + 10\log{(1+z)} + k(z)
\end{equation}

\noindent where A$_I$ is the Galactic absorption in the I band, 
$10\log(1+z)$ is the correction due to the cosmological dimming, and 
$k(z)$ is the $k$-correction.

The values for Galactic absorption adopted here are those given by
\citet{Schlegel}, and range from 0.032 to 0.363 for the clusters in the sample.
The $k$-corrections are from \citet{poggianti}, assuming her model for 
elliptical galaxies, and range from 0.006 for the  nearest cluster (A3574,
z=0.014) to 0.035 mag for the farthest one (A2670, z=0.076).  The cosmological
dimming factor ranges from 0.140 to 0.318 mag.  The median value of $\mu_t -
\mu_{lim}$ is 0.32 mag. 

The surface photometry was performed with the  task {\it ellipse}, of
IRAF/STSDAS \footnote{IRAF is distributed by the National Optical Astronomy
Observatories, which is operated by the Association of Universities for
Research in Astronomy, Inc. (AURA) under cooperative agreement with the
National Science Foundation.}. The output of this program is a list of several
parameters, such as  ellipticities, local intensities, magnitudes, etc., as a
function of  the semi-major axis value.  Using as a  radius the geometrical
average of the  semi-major and semi-minor axis, the so-called equivalent
radius, the isophotal radius ($r_{iso}$)  is estimated from the galaxy
isophotal brightness profile as the radius corresponding to $\mu_t$, and then
the isophotal magnitude ($m_{iso}$)  inside $r_{iso}$ is determined. Using the
equivalent radius instead of the semi-major axis we remove possible bias due to
different ellipticities between galaxies.

A \citet{dV} and an exponential profile were  also fitted to the isophotal
brightness profiles. Following the procedure described in \citet{SGH}, we
removed from the fitting an inner region, with radius equal to two times the
FWHM of stellar sources. These fittings were used as part of the selection of
elliptical galaxies (see Section 3.3).

The median of the formal errors are 0.02 in magnitudes  and 3.3\% in $r_{iso}$,
where errors in $r_{iso}$ have been calculated as:

\begin{equation}
\sigma (r_{iso}) = \sigma_\mu {dr \over d\mu}.
\end{equation}

\noindent with $\sigma_\mu$ being the error of the corresponding
surface brightness.
  
A check of the internal consistency of these errors can be obtained from a
sample of 16 galaxies that are present in two different images. The rms
differences between isophotal  magnitudes and isophotal radii measured in
different images are 0.02 and  2.7\%, respectively, both in agreement with the
formal errors.

\subsection{Sky subtraction} \label{ceu} Since the sky subtraction is critical
in the kind of study presented here, the procedure of sky determination is now
explained in detail. 

Firstly, the software SExtractor \citep{SEx} is used to determine  the average
sky value for the entire field.  Then the galaxy image is modeled using the
IRAF task {\it ellipse} and removed from a sky subtracted image. After, we
estimate the residuals in the local sky, by examining the residuals in a square
region centered in the subtracted galaxy, with a side size equal to at least 
two times the galaxy major-axis at $\mu_t$. The size of this region was 
determined after several tests and proved to be convenient for the estimation 
of  the local background. The new local sky level is obtained by removing the
mean value of the residuals from the previously adopted background level.

After, the procedure of surface photometry and image subtraction is done again,
using this new value for the value of the local sky. The whole process is
repeated until no significant local sky variations are found. Generally, two
iterations were enough.

This process starts with the brightest galaxy in the field and continues in
order of increasing magnitude, to avoid luminosity contamination of the galaxy
by the envelope of the BCG and other bright galaxies in the cluster field.

\subsection{Selection of the galaxy sample} \label{sample}

The selection of elliptical galaxies for the analysis was done following
several steps.

A first selection was done based on the galaxy morphology and the form of its
light profile. Initially, all galaxies that are obvious spirals were removed
from the sample.  Then, by comparing galaxy light profiles with exponential and
\citet{dV}  profiles, it was possible to remove from the sample of elliptical
galaxies some apparently featureless disk-like galaxies.

After, the morphological classification was checked  with previous
classifications  using NED \footnote{This research has made use of the
NASA/IPAC Extragalactic Database (NED) which is operated by the Jet Propulsion
Laboratory, California Institute of Technology, under contract with the
National Aeronautics and Space Administration.}.  Nearly  half  of the galaxies
in the sample had previous classifications. About 30\% of them are classified
as S0 and were removed from the sample. Only a few cases of early-type
spirals   classified as ellipticals were found. Galaxies with redshift
inconsistent with those of the clusters  were also removed.

Due to the presence of a disk component, S0 galaxies  tend to have larger
sizes, at a given magnitude, than ellipticals. However,  since the radial
distribution of S0s and ellipticals in clusters does not strictly match
each other, there is the possibility that misclassification of lenticulars
in ellipticals might produce a radial trend that could bias our results,
because the radial distribution of the former is slightly more extended than
the latter \citep{Whitmore}. This is difficult to evaluate because we do not
know, a priori, neither the number of misclassifications nor their radial
distribution. However, we expect that they would be more common in the inner
than in the outer parts of the clusters, due to the difficulty introduced by
the cluster diffuse light in the detection of faint disks. If this is indeed
the case, the bias introduced would only dilute the radial trends
discussed here.

To ensure a high photometric quality of the data,  only galaxies with 
isophotal radius larger than two times the FWHM of point sources plus 
0.8$\arcsec$ (2 pixels) were included in the sample. Additionally, images where
the chosen isophotal limit is within 1$\sigma$ of the sky brightness  were
excluded from the analysis. The faintest galaxies of the sample have 17.5 I
mag.

Objects with signs of contamination by a nearby object, which could be easily
recognized on the light profiles and in the residuals that remain after  the
subtraction of the galaxy, were also removed from the sample. Objects with
strong gradients in the local background, precluding an accurate  sky
subtraction, were also removed. 

As a final criterion, only galaxies with cluster-centric distances lower than 
R$_{200}$ were included in the analysis. This is the radius  where the mass
density is $\sim$200 times larger than  the critical density and is a good
approximation to the virial radius.  This radius R$_{200}$ is estimated using
the expression 

\begin{equation}
R_{200} = {\sqrt{3} \sigma_{cl} \over 10 H(z)}
\end{equation}

\noindent \citep{CNOC} where $\sigma_{cl}$ is the one-dimensional velocity
dispersion of  the cluster and $H(z)$, that is a function of the cosmological
parameters, is the  Hubble factor at the cluster redshift. 

The size of the sample depends of the isophotal limit chosen. Using a fiducial
isophotal limit in the galaxy rest frame of  $\mu_{lim} = 22.75$ I mag
arcsec$^{-2}$ , which avoids significant incompleteness, the full sample
contains 228 elliptical galaxies, 172 (75\%) of them with spectroscopic
confirmation of cluster membership.  The number of ellipticals 
per cluster in the final  sample is presented in Table \ref{sampletab}.

At this point it is important to warn the reader that the galaxy sample 
selected for this work is not complete in any  sense, due to the way the 
fields of the clusters were imaged and also because we have rejected a  number
of objects for which we were not able to perform the surface photometry
properly, mainly because of  light contamination by a neighbor. This problem is
more severe in the central  parts of the clusters, since there the surface
density of objects is large.  Both factors of incompleteness can change the
number of objects as a function  of the radius, but it is not expected that
they can lead to any bias that can explain the trends identified in this work 
(see next section), because they are completely independent of the  photometric
parameters of the galaxies.

\section{Results} \label{resultados}

\subsection{The magnitude-size relation for cluster ellipticals}

\begin{figure}[h!]
\centerline{\includegraphics[width=1.0\columnwidth]{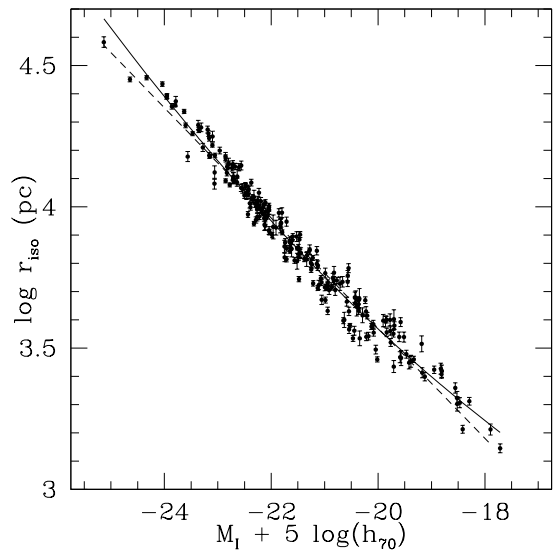}}
\vskip 0.5cm
\caption{Size-Magnitude relation in the I band
for cluster ellipticals. The isophotal limit where radii and magnitudes
are measured, in the galaxy rest frame, is 
$\mu_{lim} = 22.75$ I mag arcsec$^{-2}$. The straight dashed line
and thr continuous line are  least-squares fit to the data
using a linear and a quadratic models respectively.
\label{reta}}
\end{figure}

In Figure \ref{reta} we plot the observed relationship between the absolute
isophotal magnitude and the isophotal radius (at the fiducial isophotal limit 
in the galaxy rest frame of $\mu_{lim} = 22.75$ I  mag arcsec$^{-2}$), as well
as standard least-square fits of linear and quadratic models.

It can be appreciated in this plot that the linear model (dashed line) fits 
well the data, but with some systematic residuals in both the faint and bright
extremes. However, these residual are minimized by the use of a quadratic
model (continuous line). Actually, a likelihood ratio
test indicates that a quadratic model is significantly more reliable than a
linear model. Higher order models do not improve statistically the 
likelihood of the fit.
 
Hereafter, $<$$r_{iso}$$>$$(M)$ will be used to denote the isophotal radius of
a galaxy that, given its I magnitude, obeys the magnitude-size relation
represented by the continuos line in Figure \ref{reta}. We define
\begin{equation}
\eta \equiv {r_{iso} \over <r_{iso}>(M)},
\end{equation}
where $r_{iso}$ is the actual isophotal radius,  as a convenient way of
measuring the deviation of the isophotal radius of a galaxy from the value that
is expected given its  luminosity. Note that $\eta < 1$ represents an effective
``shrinkage'' of a galaxy.
The mean uncertainty in $\eta$  is $0.037$, taking into account
the error in the measured isophotal radius and in the fitted radius-magnitude
relation.

\begin{figure}[h!]
\centerline{\includegraphics[width=1.0\columnwidth]{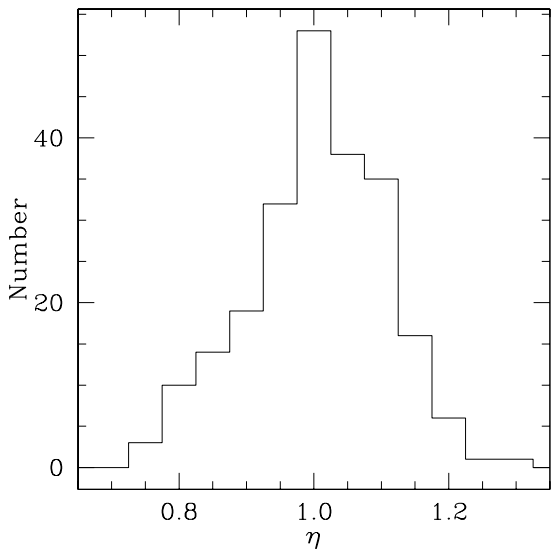}}
\vskip 0.5cm
\caption{Distribution of values of the ratio
between the measured isophotal radius of a galaxy and that expected for a
galaxy with the same magnitude, $\eta$, for the galaxies of the sample.}
\label{etahist}
\end{figure}

We present in Figure \ref{etahist} the distribution of $\eta$ for all galaxies in
our sample. Its average value is 1.005  with a standard deviation  of 0.10 (or
10\%).

The observational errors in $r_{iso}$ can account for one third
of this scatter and the errors on the magnitudes are negligible (formal errors
in $M_I$ are typically 0.02 mag). Indeed, most of the variance in $\eta$ is
due to cosmic scatter in the magnitude-size relation.

We verified that the image quality  is not affecting $\eta$  measurements. The
Spearman rank correlation coefficient between $\eta$ and  the seeing (FWHM) of
the images is 0.1, indicating  an inexistent or very low dependence between
these two quantities.

\subsection{Galaxy sizes as a function of cluster-centric distances}

In the seventies, Strom \& Strom claimed that they had found evidence for tidal
stripping in galaxy clusters based on an analysis of the isophotal radius of
cluster galaxies and a function of its cluster-centric radial position
\citep{2S}. We carry out a similar  analysis in this section using 
the cluster galaxies of our sample.

\begin{figure}[h!]
\centerline{\includegraphics[width=1.0\columnwidth]{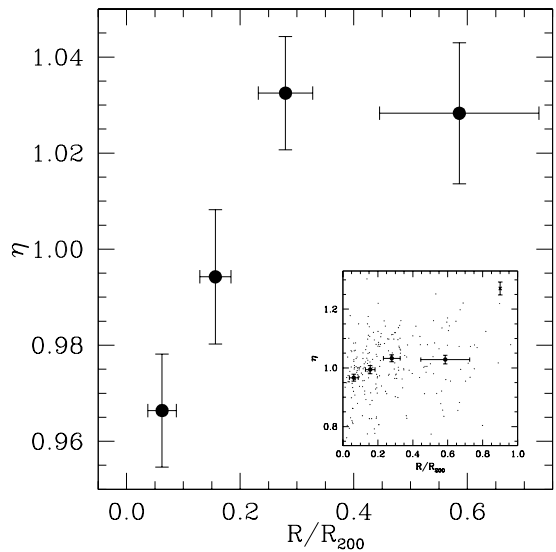}}
\vskip 0.5cm
\caption{Mean value of $\eta$ as a function of the normalized projected radial
distance $R/R_{200}$. The main plot shows bins with approximately the same
number of data points. The horizontal error bars are the 1-sigma dispersion of
the data and the vertical ones the error of the mean ($\sigma/\sqrt{N}$). The
small panel shows the individual  data points. On the upper right corner of the
small panel is shown an average $\eta$ error bar.
\label{fim}}
\end{figure}

Since the number of data  points per cluster is generally low, we have decided
to use an {\it ensemble} technique, where the data of all clusters are
analyzed  together. For that, the cluster-centric distances of the galaxies, 
$R$,  were normalized by  the value of the $R_{200}$  of the cluster.  Figure
\ref{fim} presents the ratio $\eta$ as a function of the normalized 
cluster-centric distance $R/R_{200}$ for our sample. 

To investigate radial trends in the optical extent of elliptical galaxies, we
examine whether $\eta$ varies with $R/R_{200}$. Indeed, by dividing the sample 
with respect to the median of  $R/R_{200}$ (0.202), we find that galaxies in
the group near the center ($\langle R/R_{200}\rangle$ = 0.11) have  $\eta =
0.980\pm 0.009$, whereas those in the outer group ($\langle R/R_{200}\rangle$ =
0.43) have $\eta = 1.030 \pm 0.009$. The mean difference in $\eta$ between the
two groups is (5.0$\pm$1.3)\%, ruling out at a $3.8\sigma$ (99.99\%)
confidence level (C.L.) the null hypothesis that sizes are independent of  the
cluster-centric distance.

Comparing the medians, which are more robust estimators of location than  the
means, galaxies of a given luminosity in the outer group are 5.1\% larger than
those in the inner regions of the clusters.

It is interesting to verify how this result changes as a function of
$\mu_{lim}$. In Table \ref{variacao} we show the difference in $\eta$  (for
medians and means as estimators) of galaxies in these two groups of projected
radial distances, for different isophotal limits.  For both estimators, this
difference tends to increase toward fainter values  of $\mu_{lim}$, although
this gradient is well inside the errors. 

The results of a Student's $t$-test for the statistical significance of the
difference of the mean value of $\eta$ for the two radial groups are also shown
in Table \ref{variacao} and indicates that this difference is significant  for
all values of $\mu_{lim}$. Note that the difference between the medians is 
almost always larger than that between the means.

\renewcommand{\arraystretch}{1.0}
\begin{deluxetable}{cccccc}
\tablewidth{0pt}
\tablecaption{Variation of differences of $\eta$ for galaxies in the outer and
inner radial groups, as a function of the limiting isophote\label{variacao}}
\tablehead{
\colhead{$\mu_{lim}$} & \colhead{Number} &
\multicolumn{2}{c}{Difference between} &
\multicolumn{2}{c}{Student's t-test} \\
\colhead{(mag arcsec$^{-2}$)} & \colhead{of galaxies} &
\colhead{medians (\%)}& \colhead{means (\%)} &\colhead{t} &
\colhead{Prob.}\\
\colhead{(1)} & \colhead{(2)} & \colhead{(3)}& \colhead{(4)} &\colhead{(5)} &
\colhead{(6)} \\}
\startdata
21.50 & 151 & 5.4 & 4.7 $\pm$ 1.1 & 4.4 & 100.00\%\\
21.75 & 167 & 4.9 & 5.0 $\pm$ 1.1 & 4.5 & 100.00\%\\
22.00 & 179 & 6.1 & 4.8 $\pm$ 1.2 & 4.2 & 100.00\%\\
22.25 & 191 & 5.4 & 4.6 $\pm$ 1.3 & 3.7 & ~99.97\%\\
22.50 & 210 & 4.8 & 5.0 $\pm$ 1.2 & 4.0 & ~99.99\%\\
22.75 & 228 & 5.1 & 5.0 $\pm$ 1.3 & 3.8 & ~99.98\%\\
23.00 & 200 & 8.2 & 7.0 $\pm$ 1.6 & 4.5 & 100.00\%\\
\enddata
\tablecomments{
(1) Isophotal limit in the I-band in the galaxy rest frame;
(2) Number of galaxies;
(3) Differences between the medians;
(4) Differences between the means and their errors;
(5) Value of {\it t} from the Student's {\it t}-test; 
(6) Probability that the observed difference between the two radial groups 
will not be caused by statistical fluctuations.}
\end{deluxetable}

The trend suggested by the data in Table \ref{variacao} is that the differences
between $\eta$ of the radial groups increases for fainter isophotal limits. 
This is particularly clear for the difference between medians. A word of
caution here is that the significance of this gradient is low due to the
errors in these differences.
We show in Figure \ref{fim} mean values of $\eta$ for four radial bins with
approximately the same number of data points in each one. The figure
suggests that $\eta$ increases monotonically with radius, possibly
reaching a plateau at $R/R_{200}\sim$ 0.4.

\begin{figure}[h!]
\centerline{\includegraphics[width=1.0\columnwidth]{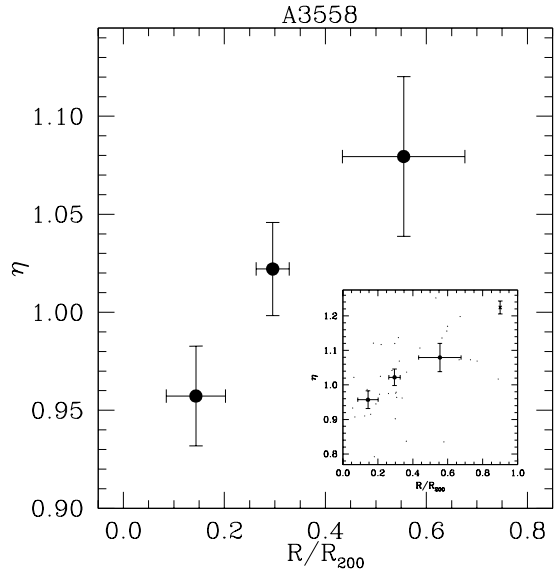}}
\vskip 0.5cm
\caption{Distribution $\eta$ as a function of the normalized projected radial
distance $R/R_{200}$ for galaxies in Abell~3558.
\label{A3558}}
\end{figure}

Unfortunately, most of the clusters do not have a sufficiently large number of 
galaxies in our sample to allow separate analyses. However, Abell~3558 is an
exception. This cluster, in the center of the Shapley super-cluster, is very
rich and has 36 galaxies in our sample, allowing a statistical analysis. The
$\eta$-radius relation for Abell~3558 is very significant, as shown in Figure
\ref{A3558}. The difference between mean values of $\eta$ below and above
$R/R_{200}=0.202$ (0.68 h$_{50}^{-1}$ Mpc) is $(9.8\pm3.6)$\%.

\renewcommand{\arraystretch}{1.0}
\begin{deluxetable}{cccccc}
\tablewidth{0pt}
\tablecaption{Variation of differences of $\eta$ for galaxies in the
outer and inner radial groups for individual clusters \label{rich}}
\tablehead{
\colhead{Cluster} & \colhead{Number} &
\multicolumn{2}{c}{Difference between}  &
\multicolumn{2}{c}{Student's t-test} \\
\colhead{ } & \colhead{of galaxies} &
\colhead{medians (\%)}& \colhead{means (\%)} &\colhead{t} &
\colhead{Prob.}\\
\colhead{(1)} & \colhead{(2)} & \colhead{(3)}& \colhead{(4)} &\colhead{(5)} &
\colhead{(6)} \\}
\startdata
A3558 & 35 & 13.6 & ~9.8 $\pm$ 3.6  &~2.5 &~98.3\%\\
A3667 & 19 & 11.1 & 10.2 $\pm$ 3.8  &~3.0 &~99.3\%\\
A496  & 19 & 14.4 & 15.7 $\pm$ 4.6  &~3.5 &~99.7\%\\
A3266 & 17 &  1.9 & ~1.6 $\pm$ 2.9  &~0.5 &~36.0\%\\
A3528 & 16 & -9.7 & -6.9 $\pm$ 6.3  &-1.2 &-75.2\%\\
\enddata
\tablecomments{Only clusters with more than 15 galaxies on the sample}
\end{deluxetable}

In Table \ref{rich} we show the difference of $\eta$ between the inner and the
outer radial groups of galaxies for the 5 clusters which have at least 15 
galaxies in our sample. These clusters are among the richest in Table
\ref{sampletab}. In four out of the five clusters the galaxies in the
outer group tend to be larger than those in the inner group, although for one
of them (A3266), the difference is not really significant. For A3528 the
difference is in the opposite direction but, again, the significance is low. 

\subsection{Galaxy sizes as a function of the local galaxy density}

We are also interested on verifying whether $\eta$ varies with the local galaxy
density. Since our imaging covers the clusters only partially, to avoid
incompleteness in the spatial sampling we have used  Super COSMOS data
\citep{super} to obtain  a catalog of the galaxies projected on to the cluster
fields.  We have selected only galaxies brighter than $R=16.5$. 

The local surface density associated to each galaxy in our sample has been
computed from the projected distance to its sixth closest Super COSMOS neighbor
(R$_6$), using the estimator of \citet{C&H}
\begin{equation}
\Sigma = {5 \over \pi R_6^2}
\end{equation}
This quantity should be corrected by contamination of foreground and background
galaxies.

This field contamination was estimated as the median  surface density of
galaxies in a ring at 4.5$\pm$0.2 R$_{200}$  of the cluster center. However,
this correction is quite sensitive to  the variance of the field galaxy counts,
often produced by large scale structures near the cluster or in its line of
sight. The cluster A3558,  in the center of the Shapley super-cluster,
illustrates well this problem.  Despite its richness (Abell richness class 4), 
the densities estimated here are among the lowest of the whole sample because
the background counts are strongly enhanced due to the presence of several
clusters in the region. For this reason, we also analyzed the data without any
background subtraction.

\begin{figure}[h!]
\centerline{\includegraphics[width=1.0\columnwidth]{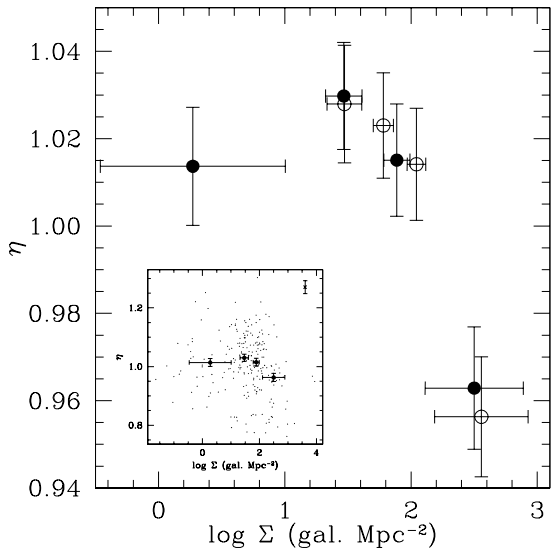}}
\vskip 0.5cm
\caption{Distribution of $\eta$ as a function of the logarithm of the projected
number density of galaxies, $\Sigma$.  The solid circles shows averages of bins
with similar number of data points for the background subtracted values of the
density. The open circles correspond to the non background-subtracted 
densities. The horizontal error bars are the 1-sigma dispersion of the
densities in each bin and the vertical ones the error of the means. In this
plot negative densities are represented as the logarithm of the absolute density
value times -1.
\label{dens}}
\end{figure}

Figure \ref{dens} shows the relation between $\eta$ and the logarithm of the
corrected projected number density. The figure suggests that there is a
tendency for $\eta$ to decrease with increasing surface density.

The difference of $\eta$ between galaxies in environments with densities larger
and smaller than the median for the background subtracted sample is 3.1\%
(median) and $3.3\pm1.3$\% (mean; 98.6\% C.L.).
The values for the non-background subtracted densities
are similar: 3.3\% (median) and $4.0\pm1.3$\% (mean; 99.7\% C.L.). The median
value of $\Sigma$  for the background subtracted sample is 51.3 Mpc$^{-2}$ and
84.3 Mpc$^{-2}$ for the non-background subtracted one. 

\subsection{Galaxy sizes as a function of cluster properties
\label{etaxcluster}}

\begin{figure}[h!]
\centerline{\includegraphics[width=1.0\columnwidth]{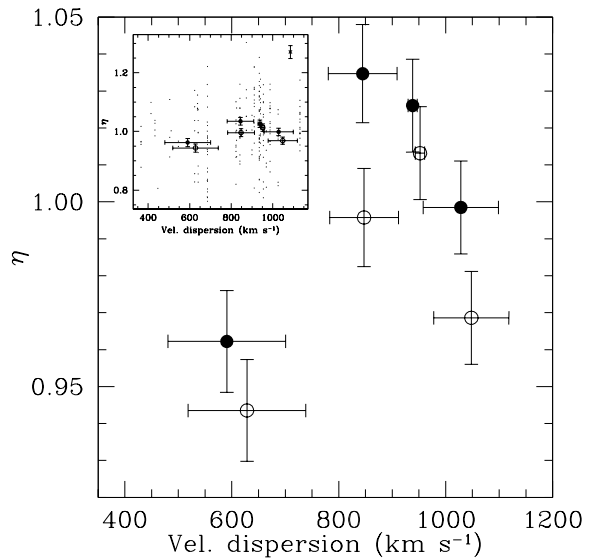}}
\vskip 0.5cm
\caption{Distribution of $\eta$ as a function of the dispersion of velocity of
the cluster. The circles represents the mean value of $\eta$ with its error of
the mean; the horizontal bars represent the interval of $\sigma_{cl}$
associated with each bin. Filled circles represent mean values for all
galaxies in the bin, whereas the open ones contain only galaxies with
cluster-centric distances less than 0.202 R$_{200}$. For clarity, we shifted
horizontally the open circles by 20 km s$^{-1}$.
\label{sigvar}}
\end{figure}

\begin{figure}[h!]
\centerline{\includegraphics[width=1.0\columnwidth]{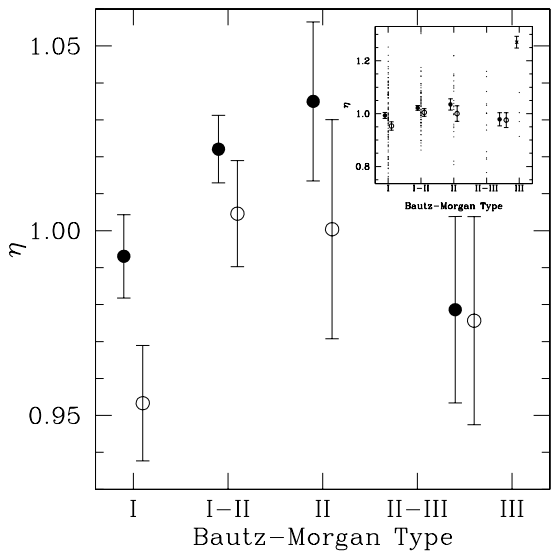}}
\vskip 0.5cm
\caption{Same as Figure \ref{sigvar}, but as a function of the 
Bautz-Morgan Type.
\label{bmvar}}
\end{figure}

In order to understand the origin of the stripping mechanism  it is important
to verify whether and how the isophotal diameters of the  galaxies depend on
cluster properties. Here we will analyze two of them: velocity dispersion and
Bautz-Morgan type that, in principle, are approximately related to mass and
evolutionary phase, respectively. In Figures \ref{sigvar} and \ref{bmvar} we
show the dependence of $\eta$ as a function of these parameters.

At a first glance neither of these figures displays a clear correlation, as
those found for the cluster-centric distance and the local density. In Figure
\ref{sigvar} it can be seen that galaxies in clusters in the smallest velocity
dispersion bin, present values of $\eta$ significantly smaller than  those in
the other clusters. 

Figure \ref{bmvar} shows how $\eta$ varies with the morphological type of the
clusters. Again we cannot see any trend here. What is noteworthy in this
figure, however, is that elliptical galaxies (excluding the BCG) in the inner
region of cD clusters (Bautz-Morgan type I) tend to be smaller than those in
other cluster types, especially those closest to the cluster center.
Actually, the inner galaxies of B-M type I clusters are the smallest in 
the present 
sample. A similar result was obtained by \citet{hardy}, who found that
elliptical and lenticular galaxies in type I clusters are dimmer, at  given
radius, than in type II clusters.

\subsection {Is there a dependence of $\eta$ with magnitude?}

\begin{figure}[h!]
\centerline{
\includegraphics[width=0.45\columnwidth]{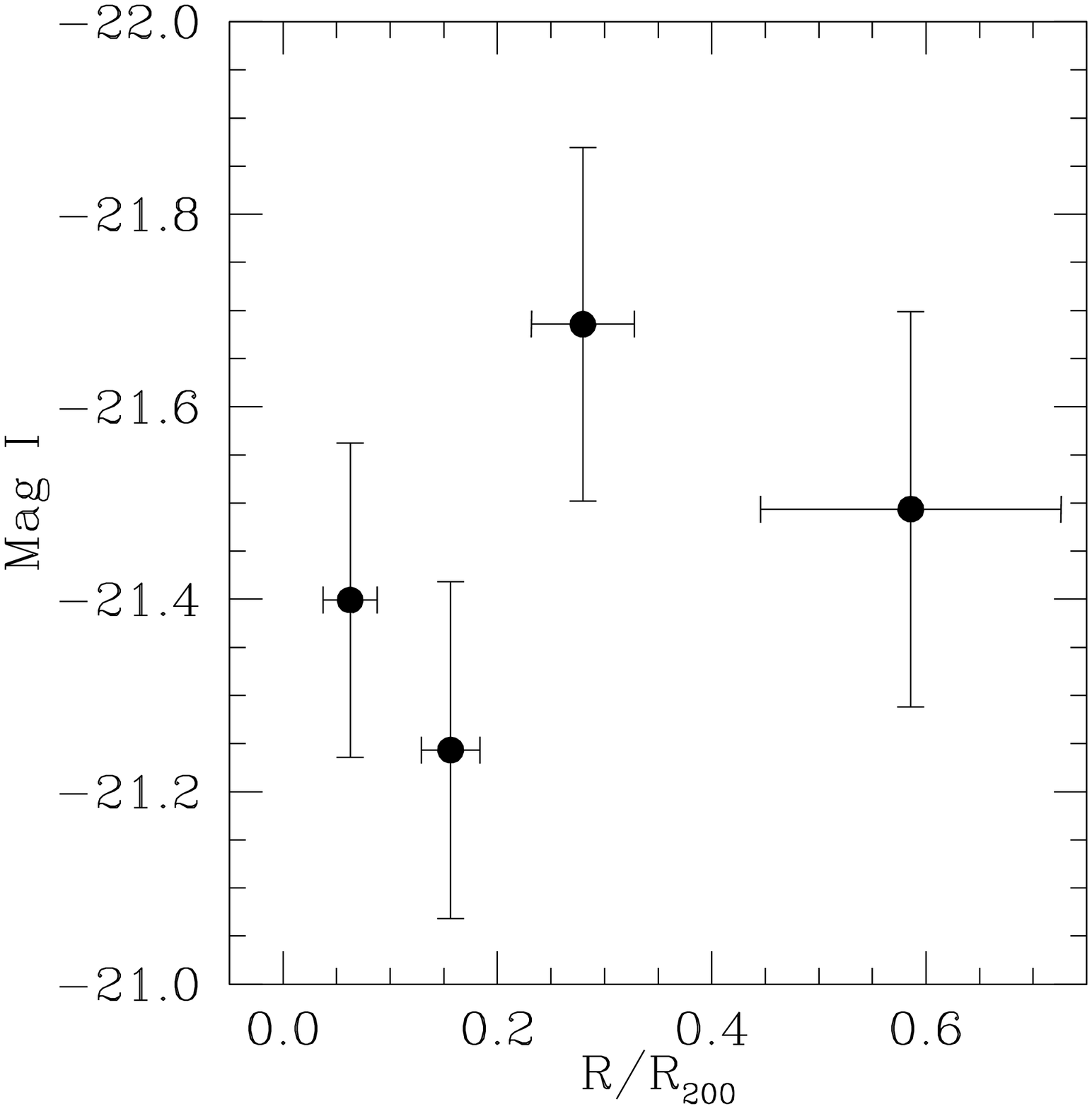} %
\includegraphics[width=0.45\columnwidth]{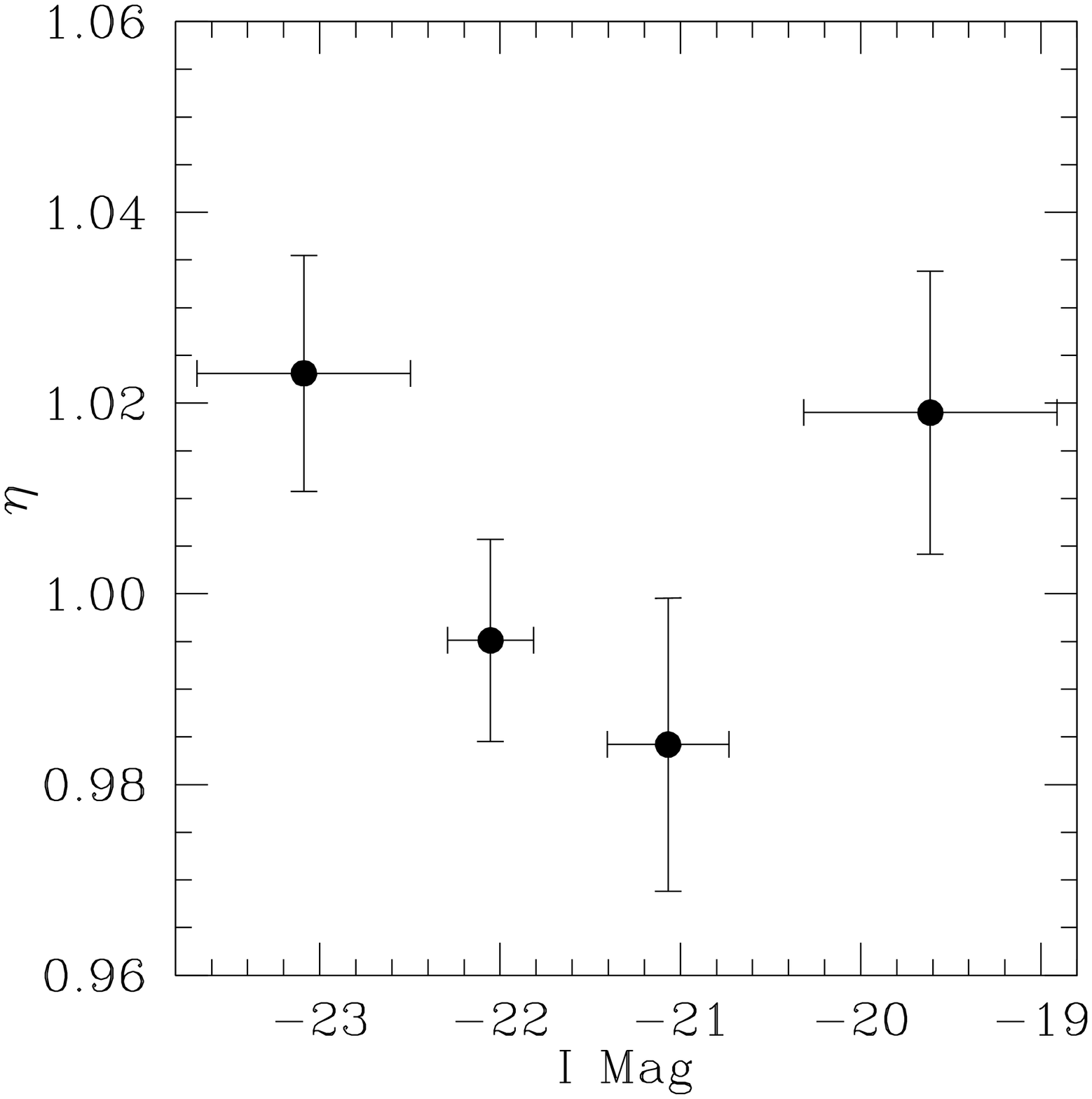}}
\vskip 0.5cm
\caption{({\it Left panel}) Mean value of the magnitude as a function of the
cluster-centric distance. 
({\it Right panel}) Mean value of $\eta$ as a function of the 
magnitude.
In both panels, the bins represent approximately the same number of
data points. The horizontal error bars are the 1-sigma dispersion of the data
and the vertical ones the error of the mean ($\sigma/\sqrt{N}$).
\label{mag}}
\end{figure}

An important check is to verify whether the effects found here are an artifact
of the fitted magnitude--size relation or the result of luminosity
segregation. We address this issue in Figure \ref{mag}.

The left panel of Figure \ref{mag}, presenting the luminosity distribution 
as a function of the cluster-centric radius, shows a hint that the
brighter galaxies tend to be found in the outer regions. The difference
considering only two radial bins, however, is rather insignificant,
$0.27\pm0.18$ mag. Moreover, the right panel reveals that there is not 
a monotonic dependence between magnitude and $\eta$.
For instance, there is no significant relative size difference between the
bright and faint half of the sample, 0.8$\pm$1.4\%. In fact, this plot
shows a difference between $\eta$ values at the extremes and the 
center of the magnitude distribution. An ispection of Figure \ref{reta}
shows that this is produced by the residuals of the fit of  the 
magnitude-size relation with our quadratic model. Rather than indicating
segregation (or anti-), the trends in Figure \ref{mag} are probably
reflecting features of the galaxy sample, which is not complete in any 
sense (c.f. Sect. 2).

For a double check, we repeated our analysis considering only 
galaxies with magnitudes between -23.5 and -20.5. Within this
interval, the linear and quadratic fits to the magnitude-size relation
are nearly identical. We obtained, by doing that, essentially the
same results, although they are noiser
due to the smaller number of data points.  We thus conclude
that all the results found in this work, regarding the relative size of
elliptical galaxies in clusters, are not artifacts of the
magnitude-size relation fit or a function of the galaxy luminosity.

\section{Discussion}

\subsection {Comparison  with previous work.}
Our main result, the shrinkage of
galaxies inhabiting the inner parts of the clusters, is in qualitative
agreement with the results of \citet{2S}. These authors found shrinkage factors
of 10-30\% in images reaching isophotal  levels as faint as 26 R mag
arcsec$^{-2}$, or $\sim$25.3 I mag arcsec$^{-2}$, assuming (R-I) = 0.7,  which
is typical of ellipticals at z$\sim$0 \citep{Fuku95}. This is nearly 2.5 mag
arcsec$^{-2}$ fainter than our  fiducial isophotal limit, 22.75 I mag
arcsec$^{-2}$, precluding a quantitative comparison between our results in
Table   \ref{variacao} with those of \citet{2S}, because the  reduction of the
isophotal radius may depend on the isophotal level.

\citet{Giuricin} have not found any environmental dependency on the sizes of 
elliptical galaxies in their sample. Their photometry reaches 25.0 B mag 
arcsec$^{-2}$, which is roughly as deep as our fiducial value for $\mu_{lim}$
\citep[assuming (B-I) $\approx$ 2.27 for ellipticals;][]{Fuku95}. However, they
warned in their conclusions that a shrinkage of  the isophotal radii of a few
percent is  consistent with their results, given their observational
uncertainties. Actually, \citet{Giuricin} report that their typical rms errors
in  $\log{r_{iso}}$ are around 0.04 (or 9.6\%), and, in the best cases  (CCD
data) they are between 0.02 to 0.03 (4.7\% to 7.2\%), well above the rms error
of the log of the isophotal radius in this work, which is 0.015 (3.5\%), as
explained in  Section 2.  Additionally, the present work uses the  ensemble
technique, that allows putting together data from different clusters, while
\citet{Giuricin} divided their sample in sub-samples with small  number of data
points, leading to results with smaller statistical  significance.

\citet{trujillo} have found a related result through the analysis of light
concentration for the Virgo and Coma clusters, indicating that more
centrally concentrated elliptical galaxies tend to
inhabit the denser regions of galaxy clusters, which is in agreement with
the results presented here. When interpreting their results, these authors
suggest that the origin of such effect would be galaxy mergers, rather than
tidal encounters because, quoting \citet{2S}, the latter would affect only the stars in the
outer halo, leaving the cores untouched. However, the numerical simulations of
\citet{AW2} convincingly demonstrate (c.f. Figure 2 of this paper) that tidal
shocks do lead to changes in the whole galaxy structure, not just in the outer
parts: ``...strong collisions produce a shrinkage in r$_e$ and a brightening of
$\mu_e$, whereas weak collisions have the opposite effect''. Here r$_e$ and
$\mu_e$ are the parameters of the de Vaucouleurs profile: the effective radius
and the surface brightness inside r$_e$.



Although tidal stripping may not be the only mechanism originating the
observed shrinkage of ellipticals, this same mechanism has been validated
through numerical simulations \citep{willman04,richstone} to explain the
observed intracluster light in clusters of galaxies. One of the findings of our
work, that, for  cD clusters (B--M type I), the shrinkage of the central
ellipticals is more pronounced than  in the other cluster types, is consistent
with the early \citet{richstone} claim that cD halos can be formed by stars
stripped out from cluster galaxies by tidal shocks, also in agreement with the
very recent work of \citep{Gonzalez2005}.

\subsection{Tidal stripping mechanism}

Finally, we discuss how can it be decided observationally whether tidal
truncation (cluster--galaxy collisions) or collisional stripping (galaxy--galaxy
encounters) is the dominant  mechanism for the shrinkage of the ellipticals.
A way to distinguish between these alternatives , is by examining how the
stripping efficiency (and thus the shrinking of a galactic radius) depends on
the cluster velocity dispersion. In the first case, as shown in equation
\ref{rmare}, the efficiency should be proportional to $\sigma_{cl}$, because
the tidal radius depends on $\sigma_{cl}^{-1}$; while, in the second case, the
dependency should be the opposite, since lower velocity dispersions leaves room
for longer lived, and hence stronger, galaxy--galaxy shocks. Such test applied
to our present data
(Section \ref{etaxcluster}), unfortunately does not show a clear correlation
between galactic sizes and $\sigma_{cl}$. 
An  analysis of a much larger sample, and preferably deeper, is necessary
to determine  which is the dominant mechanism producing the tidal
stripping of ellipticals in clusters of galaxies.

\section{Conclusions}

We have analyzed a sample of 228 elliptical galaxies belonging to 24 clusters
to look for evidence of tidal stripping and to put constraints on the
mechanisms behind star removal in cluster galaxies. The stripping was
quantified by examining departures of the magnitude-isophotal radius relation
for cluster  ellipticals, measured by the parameter $\eta$- the ratio between
the measured  isophotal radius of a galaxy (at 22.75 I mag arcsec$^{-2}$ in the
galaxy rest frame) and the isophotal radius expected for a galaxy with the same
magnitude.

Our main conclusion is that  cluster ellipticals of a given magnitude in
inner/denser cluster regions are smaller than those in the outer regions by a
factor of the order of 5\%. The shrinkage factor tend to decrease with increasing
distance from the center of the cluster and with decreasing galaxy number
density. 
Although the amount of galaxy stellar mass lost to the ICM remains
unknown, it is probably of the order of a few percent of the parent galaxy mass
within the optical radius. Thus, it is highly probable that the stars lost by
the central cluster ellipticals are a significant contributor to the diffuse
light observed in clusters of galaxies.

\acknowledgments
We would like acknowledge Gast\~ao Lima Neto, Gary Mamon, and Claudia Mendes de
Oliveira for useful discussions and Nick Suntzeff for the encouragement.
ESC acknowledges  the hospitality of the
Department of Astronomy of the IAG/USP, where most of this work has been done.
The outstanding support at Cerro Tololo Inter-American Observatory is gratefully
recognized. ESC (CNPq/Brazil--fellow) and LS acknowledges support by Brazilian
agencies FAPESP and CNPq. LEC was partially funded by Fondecyt (Chile) grant
No. 1040499.

\newpage


\end{document}